\documentstyle[12pt]{article}
\def\d{\delta}

\def\l{\lambda}

\def\o{\omega}

\newcommand{\Db}{{\bar D}}
\newcommand{\tb}{{\bar\theta}}
\newcommand{\p}[1]{(\ref{#1})}
\def\be{\begin{equation}}
\def\ee{\end{equation}}
\def\arr{\begin{array}{rll}}
\def\ea{\end{array}}
\def\bea{\begin{eqnarray}}
\def\eea{\end{eqnarray}}

\begin{document}
\renewcommand{\thefootnote}{\fnsymbol{footnote}}
\begin{titlepage}
\noindent

\vskip 3.0cm

\begin{center}

{\Large\bf New Many--Body Superconformal Models as  }\\
\bigskip

{\Large\bf Reductions of Simple Composite Systems }\\
\bigskip

\vskip 1cm

{\large Stefano Bellucci${}^{a,}$}\footnote{bellucci@lnf.infn.it},
{\large Anton Galajinsky${}^{a,b,}$}\footnote{galajin@mph.phtd.tpu.edu.ru}
and {\large Sergey Krivonos${}^{c,}$}\footnote{krivonos@thsun1.jinr.ru}

\vskip 1.0cm

{\it ${}^a$INFN--Laboratori Nazionali di Frascati, C.P. 13,
00044 Frascati, Italy}\\

\vskip 0.2cm
{\it ${}^b$Laboratory of Mathematical Physics, Tomsk Polytechnic University, \\
634050 Tomsk, Lenin Ave. 30, Russian Federation}

\vskip 0.2cm

{\it ${}^c$ Bogoliubov Laboratory of Theoretical Physics, JINR, \\
141980 Dubna, Moscow Region, Russian Federation}

\end{center}

\vskip 1cm

\begin{abstract}
We propose a new reduction mechanism which allows one to construct $n$--particle
(super)conformal theories with pairwise interaction starting from a composite
system involving $\frac{n(n-1)}{2}+1$ copies of the ordinary (super)conformal mechanics.
Applications of the scheme include an $N=4$ superconformal extension for a complexification of the
Calogero model and a $D(2,1|\alpha)$--invariant $n$--particle system.
\end{abstract}

\vspace{0.5cm}


\end{titlepage}
\renewcommand{\thefootnote}{\arabic{footnote}}
\setcounter{footnote}0

\vspace{0.5cm}

\noindent
{\bf 1. Introduction}\\

Systems describing pairwise interactions on $n$ particles in one
dimension are important for several reasons. At the classical
level they provide interesting examples of integrable models (see
Ref. \cite{per} for a review) and exhibit intriguing relations with
semisimple Lie algebras \cite{per} and Hamiltonian reductions of
$2d$ Yang--Mills theory \cite{gor}. At the quantum level they turn out
to be completely solvable (see e.g. Refs. \cite{op2},\cite{vas} and references
therein).

Apart from the mathematical concern in the chain of the
one--dimensional many--body systems, some of them admit
interesting physical applications. It has been known for a long
time that the $n$--particle Calogero model \cite{cal} characterized by the
potential $\sum_{i<j}\frac{g}{{(x_i-x_j)}^2}$, with $g$ being a
dimensionless coupling constant, exhibits conformal invariance
\cite{reg}. Exploiting the latter Gibbons and Townsend argued
recently \cite{gib} that an $N=4$ superconformal extension of the
Calogero model could be relevant for a microscopic description of
the extremal Reissner--Nordstr\"om black hole, at least near the
horizon. Unfortunately, beyond the two--particle case \cite{gal}
no such model has been constructed so far.

Generally, one works
within the framework of supersymmetric (quantum) mechanics, where
an attempt to incorporate superconformal generators leads one to
contradictory equations \cite{wyl}. Surprisingly enough, in $d=2$
the situation turns out to be simpler \cite{ghosh} and an $N=4$
superconformal Calogero model can be constructed. However, a naive
dimensional reduction to one dimension breaks $N=4$ down to $N=2$.

It is worth mentioning also the approach of Ref.~\cite{dprt}, where an
$N=4$ supersymmetric multidimensional quantum mechanics involving
an arbitrary potential has been constructed. However, no specific
choice for the potential that would reproduce the Calogero model
has been presented.

In the present paper we suggest a simple method for constructing
variant $N=4$ superconformal many--body theories
out of a composite system containing one free particle (describing the center
of mass) and $\frac{n(n-1)}{2}$ copies of the conventional $N=4$ superconformal
mechanics~\cite{ikl,ikL}. The approach appeals to a
specific reduction
where one imposes constrains on an original (simple)
model in order to generate a nontrivial potential for a resulting
reduced system. As will be discussed in detail below, the key
point is not to destroy the standard structure of a kinetic term
(i.e. $\sum_{i=1}^n {\dot x}_i {\dot x}_i$) when implementing the reduction.
As an application we construct an $N=4$ superconformal extension for a complexification of the
Calogero model and extend the $D(2,1|\alpha)$--invariant mechanics of Ref.~\cite{ikL}
to the $n$--particle case.

The paper is organized as follows. In the next section we illustrate the method and show how the
reduction works to yield the purely bosonic Calogero model. In Sect. 3
we incorporate $N=2$ supersymmetry in the scheme and reproduce the earlier
result by Freedman and Mende \cite{FM}. It turns out to be most
efficient to work off-shell and implement constraints directly in a superfield form.
Sect. 4 is split into three subsections respective
to three different $N=4$ off--shell superconformal multiplets known.
Here we construct the two models mentioned above and comment on the
problems one reveals when trying to apply the method for constructing an
$N=4$ superconformal extension for the real Calogero model.
Possible further developments are outlined in the concluding Sect. 5.

\noindent
{\bf 2. The bosonic Calogero model}

\vspace{0.5cm}

Before we treat in detail the entire superconformal case, it is worth illustrating how
the reduction works for bosonic systems, the
simplest conformally invariant many--body theory being the three--particle Calogero model in one
dimension. To this end, let us consider the composite theory which involves a free particle and three copies of
conformal mechanics~\cite{aff} with the (dimensionless positive) coupling constants being equal to each other
\bea\label{start}
&&
S_0=\int dt \left[ \frac{{\dot x}_0^2}{2} +\sum_{i=1}^3(\frac{{\dot x}_i^2}{2} -\frac{g}{x_i^2})  \right].
\eea
The free particle degree of freedom has a distinguished status and is reserved for the center of
mass coordinate of a resulting many--body conformal model to be derived below. Being a composite system, the
theory~(\ref{start}) possesses an invariance which is the direct product of four $SO(1,2)$ groups. Aiming at the construction of
a many--body interacting theory exhibiting a {\it single} conformal symmetry, let us reduce the model~(\ref{start})
to the plane in the configuration space at hand
\be\label{plane}
x_1+x_2+x_3=0.
\ee
This constraint can be readily enforced by means of the Lagrange multiplier $\pi$, the reduced action functional
acquiring the form
\bea\label{act}
&&
S=\int dt \left[ \frac{{\dot x}_0^2}{2} +\sum_{i=1}^3(\frac{{\dot x}_i^2}{2} -\frac{g}{x_i^2})+
\pi(x_1+x_2+x_3)  \right].
\eea

We next wonder which is the dynamics of physical degrees of freedom in the reduced theory.
Applying the standard Dirac method one finds\footnote{$(p_0,p_i,p_\pi)$ denote momenta canonically
conjugate to the configuration space variables $(x_0,x_i,\pi)$.}
\be\label{con1}
p_\pi=0,
\ee
as the primary constraint and Eq.~(\ref{plane}) as the secondary one. Besides, there appear two tertiary constraints
\be\label{con2}
p_1+p_2+p_3=0, \quad \pi+\sum_{i=1}^3 \frac{2g}{3x_i^3}=0,
\ee
whose conservation in time fixes the Lagrange multiplier $\l_\pi$ entering the
canonical Hamiltonian
\be\label{hamil}
H=\frac{p_0^2}{2}+\sum_{i=1}^3 \frac{p_i^2}{2}+p_\pi \l_\pi +\sum_{i=1}^3 \frac{g}{x_i^2}-\pi (x_1+x_2+x_3).
\ee
Its explicit form is irrelevant for the subsequent consideration.

It is important to notice that only a specific linear combination of the original
conserved charges
\bea
&&
D=tH-\frac{1}{2} \sum_{i=0}^3 x_n p_n -p_\pi \sum_{i=1}^3 \frac{g}{x_i^3},
\quad
K=t^2 H -t(\sum_{i=0}^3 x_n p_n +2p_\pi \sum_{i=1}^3 \frac{g}{x_i^3})+\frac{1}{2} \sum_{i=0}^3 x_n x_n,
\nonumber\\[2pt]
&&
\eea
leaves the constraint surface~(\ref{plane}),(\ref{con1}),(\ref{con2}) invariant\footnote{There is also an extra $SO(1,2)$ invariance
realized on the center of mass coordinates $(x_0,p_0)$ which we ignore here.}. Together with the
Hamiltonian~(\ref{hamil}) they form the conformal algebra $so(1,2)$ modulo the constraint terms.

Alternatively, working in the Lagrangian framework one could observe that the coordinates of
each individual particle entering the original composite system transform
{\it linearly} and {\it homogeneously} with respect to the proper $SO(1,2)$ group
\be\label{cm4}
\delta x_i =\frac{1}{2} \dot f  x_i, \quad  \delta t = a+ bt+ct^2 \equiv f(t).
\ee
In fact, it is this observation which prompts one to impose the linear constraint (\ref{plane})
and guarantees the residual "diagonal" conformal symmetry for the reduced system.

Being second class, the constraints revealed above can be resolved
after introducing the Dirac bracket. This allows one to remove the pair $(\pi,p_\pi)$ and, say $(x_3,p_3)$, from the consideration,
while brackets for the remaining variables prove to have the following form (only
the non vanishing brackets are given):
\bea\label{bracket}
&&
{\{ x_0,p_0 \}}_D=1, \quad {\{x_1,p_1 \}}_D=\frac{2}{3}, \quad {\{x_1,p_2 \}}_D=-\frac{1}{3}, \quad
{\{x_2,p_1 \}}_D=-\frac{1}{3}, \quad {\{x_2,p_2 \}}_D=\frac{2}{3}.
\nonumber\\[2pt]
&&
\eea
It is then easy to find an invertible change of variables $(x_0,x_1,x_2) \rightarrow (q_1,q_2,q_3)$ which
diagonalizes the bracket~(\ref{bracket})
\bea
&&
x_0=q_1+q_2+q_3, \quad x_1=q_1-q_2, \quad x_2=q_2-q_3,
\nonumber\\[2pt]
&&
p_0=\frac{1}{3}(p_{q_1}+p_{q_2}+p_{q_3}), \quad p_1=\frac{1}{3}(p_{q_1}-p_{q_2}),
\quad p_2=\frac{1}{3}(p_{q_2}-p_{q_3}),
\eea
with ${\{q_i,p_{q_j} \}}_D=\d_{ij}$. Being rewritten in terms of the new variables, the equations of
motion coincide with those following from the Hamiltonian\footnote{Here we made the additional trivial change
$p'_{q_i}=1/\sqrt{3} p_{q_i}$, $q'_i=\sqrt{3} q_i$ and $g'=3g$.}
\be\label{calogero}
H=\frac{1}{2} \sum_{i=1}^3~ p'_{q_i}{}^2  +\sum_{i<j}~ \frac{g'}{{(q'_i-q'_j)}^2},
\ee
which is precisely the Hamiltonian of the three--particle Calogero model.

Thus, starting from
the simple system composed of a free particle (the center of mass) and three copies of the conformal mechanics and applying
the reduction~(\ref{plane}) one produces the three--particle Calogero model
with the pairwise interaction in one dimension.

Our analysis above relies upon the rigorous Hamiltonian procedure. However, the same conclusion can be reached in a
simpler way if one just solves the constraint~(\ref{plane})
\be
x_1=q_1-q_2, \quad x_2=q_2-q_3, \quad x_3=q_3-q_1,
\ee
immediately in the Lagrangian~(\ref{start}), with $x_0$ being the center of mass, $x_0=q_1+q_2+q_3$.
It is worth mentioning that a peculiar feature of the reduction used is that it preserves the kinetic term, i.e.
$\sum_{k=0}^4 {\dot x_k}^2=3 \sum_{i=1}^3 {\dot q_i}^2$.

Obviously, the method is general enough and is applicable to any $n$--particle system with a potential of the form
$\sum_{i<j} V(q_i-q_j)$. For example, the $n$--particle Calogero model with the harmonic oscillator potential
\be\label{harmcal}
S=\int dt \left(\sum_{i=1}^n \frac{{\dot q}_i^2}{2} -\sum_{i<j} \o^2 {(q_i-q_j)}^2
-\sum_{i<j}\frac{g}{{(q_i-q_j)}^2}\right),
\ee
can be obtained from a composite system which contains a free particle and $\frac{n(n-1)}{2}$ copies of the conformal mechanics
which are modified by the inclusion of the harmonic oscillator potential
\be
S=\int dt \left(\frac{{{\dot x}_0}^2}{2} +  \sum_{i<j}^{n}( \frac{{\dot x_{ij}}^2}{2}  -\frac{g}{x_{ij}^2} -
 \o^2 x_{ij}^2) \right)\quad , \quad x_{ij}=-x_{ji}\;.
\ee
The aforementioned  reduction is implemented with the help of the relations
\be\label{red}
x_0= \sum_i q_i \; , \quad x_{ij} = q_i-q_j, \; i<j \quad .
\ee
The variables $x$ obey $(n-1)(\frac{n}{2}-1)$ linear constraints which leave
one with $n$ physical degrees of freedom $(q_1,\dots,q_n)$ and the Calogero
action functional~(\ref{harmcal}).

The fact that one can treat the $n$--particle Calogero model as a specific  reduction of a simpler
composite system is not just the matter of aesthetics. It is well known that conformal transformations
characterizing the model~\cite{aff} are compatible with $N=2$ sypersymmetry~\cite{ap} and
$N=4$ supersymmetry~\cite{FR,ikl}.
The trick described above suggests quite new and intriguing possibility to
attack the problem of a superconformal generalization of the $n$--particle Calogero model. It suffices to
start with a composite system involving a superconformal extension of a free particle and $\frac{n(n-1)}{2}$
copies of the superconformal mechanics~\cite{ikl} and apply an appropriate reduction.
In the next section we treat in detail the $N=2$ case. $N=4$ models are discussed in Sect. 4.

\vspace{0.5cm}

\vspace{0.5cm}

\noindent
{\bf 3. $N=2$ superconformal  Calogero model}

\vspace{0.5cm}

We next wonder how to incorporate $N=2$ supersymmetry in the reduction scheme outlined above.
In analogy with the bosonic case, in order to guarantee a residual superconformal symmetry for a
reduced system, one is to impose constraints on the coordinates which transform linearly and
homogeneously under the superconformal group. This suggests, in particular,
that one should work with superfields because, in general, after eliminating
auxiliary fields supersymmetry transformations become nonlinear.

Thus, our starting point is the $N=2$ superconformal mechanics \cite{ap,ikl} which is described by the
superfield action functional
\be\label{n2scm}
S_0=\frac{1}{2}\int dt d\theta d\bar\theta \left( D X \Db X -2 g \log | X | \right) \;.
\ee
Here $X(t,\theta,\tb)$ is a bosonic $N=2$ real superfield and the covariant derivatives are defined
in the usual way
$D=\frac{\partial}{\partial \theta}+i\tb \frac{\partial}{\partial t}$,
$\Db=\frac{\partial}{\partial \tb}+i\theta \frac{\partial}{\partial t}$,
$\left\{ D,\Db\right\}=2i\frac{\partial}{\partial t}$.

The action \p{n2scm} holds invariant with respect to the following transformations
\be\label{n2tr}
\delta t=E-\frac{1}{2}\tb\Db E-\frac{1}{2}\theta D E\;,\;\delta\theta=-\frac{i}{2}\Db E\;,\;
\delta\tb = -\frac{i}{2} D E\;,\;\delta X= \frac{1}{2}{\dot E}X\;,
\ee
where the superfunction $E(t,\theta,\tb )$ collects all the infinitesimal parameters of the $N=2,d=1$
superconformal group
\be
E(t,\theta,\bar\theta ) = f(t)-2i\left(\varepsilon+\beta t\right) \tb -2i\left(\bar\varepsilon + \bar\beta t\right)\theta
+\theta\tb h \;.
\ee
Here $f(t)$ is the same as that in Eq. (\ref{cm4}) above, $h$ is a $U(1)$ rotation parameter, while
$\varepsilon$ and $\beta$ correspond to the Poincar\'e and conformal supersymmetries, respectively.

Aiming at the construction of an $N=2$ superconformal Calogero model, let us consider a composite
system which involves an $N=2$ superconformal extension of a free particle
and $n(n-1)/2$ copies of the $N=2$ superconformal mechanics discussed above
\be\label{n2cal1}
S=\frac{1}{2}\int dt d\theta d\bar\theta\left[ D X_0\Db X_0 +\sum_{i<j} \left( D X_{ij} \Db X_{ij} -2 g \log |X_{ij}| \right)\right] \;,
\ee
where
$X_{ij}=-X_{ji}$ and $i,j =1,2,\ldots , n$.
Alike the bosonic case, we then impose the reduction constraints which express $n(n-1)/2+1$ original
superfields $X_0,X_{ij}$ in terms of $n$ new superfields $V_i$
\be\label{n2hamred}
X_0=\frac{1}{\sqrt{n}}\sum_i V_i \;\quad X_{ij}=\frac{1}{\sqrt{n}} \left( V_i-V_j \right) \;.
\ee
As each of the superfields $X_0,X_{ij}$ transforms linearly and homogeneously
with respect to the proper $SU(1,1|1)$ transformation, the constraints \p{n2hamred}
hold invariant under the residual $SU(1,1|1)$ group acting in a uniform way on all the superfields
\be\label{n2tr1}
\delta V_i =\frac{1}{2}{\dot E} V_i \;.
\ee
Thus, a residual $N=2$ superconformal symmetry is guaranteed by the construction.

Implementing the reduction directly in the action functional of the composite system
\p{n2cal1} one gets the following $N=2$ superconformal $n$-particle model with the pairwise interaction
\be\label{n2cal2}
S=\frac{1}{2}\int dt d\theta d\bar\theta \left( \sum_{i=1}^n D V_{i} \Db V_{i} -2 g \sum_{i<j} \log |V_i-V_j| \right) .
\ee
Eliminating the auxiliary fields $F_i=\left[\Db,D\right]V_i|_{\theta=0}$ with the use of
their equations of motion one can formulate the model in terms of the
physical degrees of freedom
\be\label{n2cal3}
S=\frac{1}{2}\int dt \left[ \sum_{i=1}^n ({\dot{q_{i}}}^2 -i\bar\psi_i{\dot\psi_i} +
 i\dot{\bar\psi_i}\psi_i) - \sum_{i<j} \frac{g^2-g(\psi_i-\psi_j)(\bar\psi_i-\bar\psi_j)}{(q_i-q_j)^2} \right] ,
\ee
where
$q_i={V_i}|_{\theta=0}$, $\psi_i=iD {V_i}|_{\theta=0}$ and ${\bar\psi}_i=i\Db {V_i}|_{\theta=0}$.
Being rewritten in the Hamiltonian form, the latter theory coincides with the $N=2$ superconformal
Calogero model constructed by Freedman and Mende in the framework of many--body
supersymmetric quantum mechanics \cite{FM}.

Thus, we indeed checked that the reduction scheme advocated in the previous section can incorporate the $N=2$ supersymmetry.
Obviously, one can employ this reduction, in order to build supersymmetric many--body systems with a potential of
a more general form $\sum_{i<j} \Phi(V_i-V_j)$. It suffices to start with a proper composite
system.

\vspace{0.5cm}

\noindent
{\bf 4. Calogero--type models with $N=4$ superconformal symmetry}

\vspace{0.5cm}
The situation becomes more complicated when one tries to extend the analysis, in order to incorporate
$N=4$ supersymmetry. Let us remind first that the $N=4$ superconformal group in $d=1$ is the exceptional
supergroup $D(2,1|\alpha)$ \cite{FrSorba}. Only for specific values of the parameter, namely $\alpha=-1,0$ (with the notation
from Ref. \cite{bils}), the corresponding superalgebra
is isomorphic to $su(1,1|2)\oplus su(2)$. Generally, only the $su(1,1|2)$ part is taken into account,
while the $su(2)$ subalgebra is kept explicitly broken. Besides, three different
$N=4$ off-shell supermultiplets are known, which can be used for constructing $N=4$ superconformal Calogero--type models.
The first multiplet contains one real scalar, four real fermions and three real auxiliary fields \cite{ikl}.
We shall call this the {\it 4a} multiplet. The second multiplet involves
one complex scalar, four real fermions and one complex auxiliary scalar \cite{ap,FR}. We will call this the {\it 4b} multiplet.
Finally, it was recognized recently that
a multiplet with three real scalars, four real fermions and one real auxiliary field \cite{IS,BP,MSS} underlies
a new version \cite{ikL} of an $N=4$ superconformal mechanics (for the case of a vanishing coupling constant,
see also Ref.~\cite{bgik}). We will call the latter the $4c$ multiplet.
It is important to stress that in all three cases the
$N=4$ superfields involved are subject to constraints. Therefore, one should check that the proposed reduction is
compatible with those conditions.

Keeping all this in mind, we proceed to construct
$N=4$ superconformal Calogero--type models.

\vspace{0.5cm}

\noindent
{\bf 4.1 The {\it 4b} multiplet}

\vspace{0.5cm}

The {\it 4b} multiplet proves to be the simplest multiplet to handle. It underlies
the following $N=4$ superconformal mechanics \cite{ap,FR,ikl}
\be\label{4baction}
S=\frac{1}{2} \left( \int dt d^4\theta Y{\bar Y} -g\int dt d^2\theta \log |Y| -g \int dt d^2 \tb \log |{\bar Y}|\right) \;,
\ee
where the $N=4$ superfields $Y,{\bar Y}$ are constrained to obey the chirality conditions
$\Db^a Y=0$, $D_a {\bar Y}=0\;$
which involve the covariant derivatives
$D_a=\frac{\partial}{\partial \theta^a}+i\tb_a\frac{\partial}{\partial t}$, $
\Db^a=\frac{\partial}{\partial \tb_a}-i\theta^a \frac{\partial}{\partial t}$, $a=1,2$.

The $SU(1,1|2)$ symmetry characterizing the theory is realized in the following way
\be\label{4btr}
\delta t=E-\frac{1}{2}\theta^2 D_a E+\frac{1}{2}\tb_a\Db^a E\;,\;
\delta \theta^a=\frac{i}{2}\Db^a E\;,\; \delta\tb_a=-\frac{i}{2}D_a E\;,\quad \delta Y =\dot{E_L}(t_L,\theta)Y\;,
\ee
where the superfunction $E=f-2i\left( \varepsilon(t)\tb -\theta\bar\varepsilon (t)\right) +\frac{1}{2}\left( \theta \tau^k \tb\right) b^k -2
 \left( \dot\varepsilon \tb +\theta\dot{\bar\varepsilon}\right)\theta\tb+\frac{1}{2}\left(\theta\tb\right)^2 \ddot{f}$
collects all the parameters of the $SU(1,1|2)$ group. Here we made use of the notation
$f=a+bt+ct^2$, $\varepsilon^a (t) =\varepsilon^a+\beta^a t$,
$E_L=\frac{1}{2}f(t_L)+2i\theta^a{\bar\varepsilon}_a(t_L)$, $t_L=t+i\theta \bar\theta$ and $t_R=t-i\theta \bar\theta$.

Owing to the homogeneous transformation law \p{4btr} for $Y$ and ${\bar Y}$, one can safely use this model as a constituent of
a larger composite system. As before, the latter is taken to be the sum of a kinetic term for the superfield
$Y_0$ $(g=0)$ and $n(n-1)/2$ extra terms of the form \p{4baction} with $Y_{ij}=-Y_{ji}$.
Imposing the reduction constraints on such a composite system
\be\label{n4hamred}
Y_0=\frac{1}{\sqrt{n}}\sum_i V_i \;\quad Y_{ij}=\frac{1}{\sqrt{n}} \left( V_i-V_j \right) \;,
\ee
where  $V_i$ denote $n$ new independent superfields obeying
the same chirality conditions $\Db^a V_i=0$, $D_a {\bar V_i}=0$, one ends up with
the superfield action functional
\be\label{4bCal}
S_{4b}=\frac{1}{2} \left( \int dt d^4\theta  \sum_i V_i{\bar V_i}
-g\int dt d^2\theta \sum_{i < j}\log |V_i-V_j| -
g \int dt d^2 \tb \sum_{ i < j} \log | {\bar V_i} -{\bar V_j}| \right) \;.
\ee
By construction this holds invariant under the "diagonal" $SU(1,1|2)$ transformations
which act in a uniform way on all the superfields $V_i,{\bar V_i}$.

In order to clarify the status of this model, we eliminate auxiliary fields and go over to
physical components
\bea\label{4bCalcomp}
S_{4b}=&&\frac{1}{2}\int dt\left[ \sum_i \left( {\dot q_i}{\dot {\bar q_i}} +\frac{1}{2}{\dot\chi_i}{\bar\chi_i}\right)-
\sum_{i<j}\left(  \frac{g^2}{(q_i-q_j)(\bar q_i -\bar q_j)} -\frac{g (\chi_i-\chi_j)(\chi_i-\chi_j)}{4(q_i-q_j)^2} -\right.\right.
\nonumber \\
&& \left. \left. -\frac{g(\bar\chi_i-\bar\chi_j)(\bar\chi_i-\bar\chi_j)}{4(\bar q_i -\bar q_j)^2}
\right) \right] \;,
\eea
where $q_i={V_i}|_{\theta=0}$, $\bar{q_i} ={\bar V_i}|_{\theta=0}$, $\chi_{ia}=-i D_a {V_i}|_{\theta=0}$,
${\bar\chi}_i^a=i\Db^a {\bar V_i}|_{\theta=0}$. As is seen from the component action, the theory at hand
describes an $N=4$ superconformal extension of a complexification of the $n$--particle Calogero model.
It is worth noting that taking a real slice of the latter theory one breaks the $N=4$ supersymmetry and, hence,
an $N=4$ superconformal {\it real} Calogero model can not be constructed in this way.
We will return to this issue later on in Sect. 4.3.

\vspace{0.5cm}

\noindent
{\bf 4.2 The {\it 4c} multiplet}

\vspace{0.5cm}

This model is based on a recently proposed version \cite{ikL} of
an $N=4$ superconformal mechanics. In the bosonic sector the
system contains a dilaton field together with two variables
parameterizing the two-sphere $S^2\sim SU(2)/U(1)$. It was
suggested in \cite{ikL} that this variant of the superconformal
mechanics is a disguised form of a charged superparticle
moving in $AdS_2\times S^2$ background. Inspired by the
use of the $n$--particle Calogero model in the
context of black hole physics \cite{gib}, in this section we extend the system to
the $n$-particle case.

The starting point is the superfield action functional
\be\label{4cactionA}
S=-\int dt d^2\theta\left[ \frac{1}{4}\left(
v^2+4\rho\bar\rho\right)^{\frac{1-2\alpha}{2\alpha}}\left( D v \Db
v + D\rho \Db{\bar \rho}\right) +
 g \log \left| \frac{v+\sqrt{v^2+4\rho\bar\rho}}{2}\right| \right]
 \;,
\ee which depends on the parameter $\alpha$ characterizing the
supergroup $D(2,1|\alpha)$, the latter being the symmetry group of the model.
The superfield $v$ entering the problem is an unconstrained bosonic $N=2$ superfield,
while the $N=2$ superfields $\rho$ and ${\bar \rho}$ are taken to obey the chirality
conditions $D \bar\rho=0$, $\Db \rho =0.$ Altogether they form
an irreducible $N=4$ supermultiplet \cite{IS,BP,MSS}.

The fact that $v,\rho,\bar\rho$ transform linearly and homogeneously under the action of the $D(2,1|\alpha)$
supergroup allows one to use this theory for the construction of a composite system and then apply to the latter
the reduction scheme. Omitting the details we expose the resulting action functional
\bea\label{4cCalg}
S&=&-\int dt
d^2\theta\left[ \frac{1}{4}\left( Y_0^2+4 X_0\bar
X_0\right)^{\frac{1-2\alpha}{2\alpha}}\left( D Y_0 \Db Y_0 + D X_0
\Db{\bar X_0}\right) +
     \right.  \\
&& \left. \frac{1}{4}\sum_{i<j} \left( Y_{ij}^2+4 X_{ij}\bar X_{ij}\right)^{\frac{1-2\alpha}{2\alpha}}
\left( D Y_{ij} \Db Y_{ij} + D X_{ij} \Db{\bar X_{ij}}\right)+
 g \log \left| \frac{Y_{ij}+\sqrt{Y_{ij}^2+4X_{ij}\bar X_{ij}}}{2}\right| \right] \;, \nonumber
\eea
where $Y_0=\frac{1}{\sqrt{n}}\sum_{i=1}^n
v_i$, $X_0=\frac{1}{\sqrt{n}}\sum_{i=1}^n \rho_i$, $Y_{ij}=\frac{1}{\sqrt{n}} \left(
v_i-v_j\right)$, $X_{ij}=\frac{1}{\sqrt{n}}\left(
\rho_i-\rho_j\right)$ and the superfields
$\bar\rho_i(\rho_i)$ are chiral (anti-chiral).

As formulated in Eq. \p{4cCalg} above, the action functional exists
for generic values of the parameter $\alpha$ {}\footnote{A singular
value of the parameter $\alpha=0$ is equivalent to $\alpha=-1$,
see e.g. Ref. \cite{ikL}.}. Notice, however, that only for $\alpha=1/2$ the kinetic
term takes the standard flat form. In what
follows we concentrate on this particularly interesting case for which
the action functional is simplified drastically. After eliminating
auxiliary fields with the use of their equations of motion
one finds the resulting component action
\bea
S& =& \int dt \sum_i\left( \frac{1}{2} {\dot v_i}^2 +2 {\dot \rho_i}{\dot{\bar\rho_i}} -\frac{i}{2}\dot\xi_i\bar\xi_i+
\frac{i}{2}\xi_i\dot{\bar\xi_i}-\frac{i}{2}\dot\psi_i\bar\psi_i+\frac{i}{2}\psi_i\dot{\bar\psi_i} \right) - \nonumber \\
&& \sum_{i<j} \left(    \frac{8g^2}{v_{ij}^2+4\rho_{ij}{\bar\rho_{ij}}}+
 \frac{8ig \left( \rho_{ij} \dot{\bar\rho_{ij}}-\dot\rho_{ij}\bar\rho_{ij} \right) }
 { \sqrt{v_{ij}^2+4\rho_{ij}{\bar\rho_{ij}}} \left( v_{ij}+\sqrt{ v_{ij}^2+4\rho_{ij}{\bar\rho_{ij}}}\right)} \right.+ \nonumber \\
&& \left. \frac{4g}{(v_{ij}^2+4\rho_{ij}{\bar\rho_{ij}})^{3/2}}\left( v_{ij}\xi_{ij}{\bar\xi_{ij}} -v_{ij}\psi_{ij}{\bar\psi_{ij}}+
   2\rho_{ij}\xi_{ij}{\bar\psi_{ij}}+2{\bar\rho_{ij}\psi_{ij}{\bar\xi_{ij}}} \right) \right)\;,
 \eea
where the physical components are defined as follows
$v_i=v_i|_{\theta=0}$, $\rho_i=\rho_i|_{\theta=0}$, $\xi_i=Dv_i|_{\theta=0}$, $\psi_i=D\rho_i|_{\theta=0}$
and we made use of the notation
$\rho_{ij} = \rho_i-\rho_j$, $v_{ij}=v_i-v_j$, $\psi_{ij} =\psi_i-\psi_j$, $\xi_{ij}=\xi_i-\xi_j$.

To summarize, the model constructed in this subsection can be viewed as an $N=4$
superconformal Calogero--type model in three dimensions.

\newpage

\noindent {\bf 4.3 Troubles with the {\it 4a} multiplet}

\vspace{0.5cm}

Being the most interesting multiplet in view of the conjecture of
Ref. \cite{gib}, this case turns out to be the most difficult to
handle. In the framework of the Hamiltonian formalism the trouble
has been revealed already in Ref. \cite{wyl} (see also the related
work \cite{gal}). In order to encounter the problem
in our setting, let us recall the form of the action functional of an $N=4$
superconformal mechanics which relies upon the multiplet at hand~\cite{ikl, ikp}
\be\label{4a}
S=\frac{1}{16}\int dt d^4 \theta \left( e^u
-\frac{1}{8} \theta^a_\alpha \theta_{\beta a} m^{\alpha\beta} u
\right).
\ee
Here $m^{\alpha\beta}$ is a constant $SU(2)$ vector
which plays the role of a coupling constant, whereas the $N=4$ superfield $u$
is constrained to obey the equation
\be\label{2acon}
D_{a(\alpha}D^\alpha_{b)}u=0 \;.
\ee
Because the superfield $u$
transforms linearly and homogeneously with respect to the supergroup
$SU(1,1|2)$ one can still try to use this theory in the context of the procedure
we developed in the preceeding sections. As
an outcome of this one finds the action functional
\be\label{4aCal}
S=\frac{1}{16}\int dt d^4 \theta \left[ e^{Y_0}
+\sum_{i<j}\left( e^{Y_{ij}} -\frac{1}{8} \theta^a_\alpha
\theta_{\beta a} m^{\alpha\beta} Y_{ij}\right) \right],
\ee
where, as before,
$Y_0=\frac{1}{\sqrt{n}}\sum_{i=1}^n v_i$, $Y_{ij}=\frac{1}{\sqrt{n}} \left( v_i-v_j\right)$
and all the superfields $v_i$ satisfy the constraints \p{2acon}. A simple
inspection then shows that the kinetic term of the component action
includes a non-trivial coupling to a specific  metric and
thus the model fails to reproduce the real Calogero model in the
bosonic limit. A way out could be to perform the reduction not in
terms of the superfield $u$, but rather $e^{\frac{1}{2}u}$. In
other words, one can pass to the new variable
$w=e^{\frac{1}{2}u}$ in the action \p{4a} and then apply the
same reduction as before for the newly introduced superfields.
Unfortunately, this step proves to be incompatible with the
constraints \p{2acon} which are of the second order in the covariant
derivatives. Thus, for the {\it 4a} multiplet our reduction
fails to produce an $N=4$ superconformal extension of the real
Calogero model.

\vspace{0.5cm}

\noindent {\bf 5. Concluding remarks }

\vspace{0.5cm}

To summarize, in this paper we implemented a new reduction mechanism and
constructed Calogero--type models with extended superconformal symmetry starting
from simple composite systems.
As was mentioned above, the range of possible applications of the scheme
is actually larger than just the (super) conformal theories.
Whenever the constraints preserve the standard structure of a kinetic term, a many--body
system with the general potential $\sum_{i<j}V(x_i-x_j)$ can be constructed.
Using this variant of the reduction  we built
an $N=4$ superconformal extension for a complexification of the
Calogero model and a $D(2,1|\alpha)$--invariant $n$--particle system.

Turning to possible further developments it seems interesting to study
whether the reduction proposed in this work allows one to generate
solutions for a reduced $n$--particle model proceeding from those characterizing
an initial simple composite system. A related question is whether the
$D(2,1|\alpha)$--invariant many--body theory constructed in this paper
is integrable. It is also interesting to examine the range of applicability of this
scheme in the context of two-dimensional field theory.

\vspace{0.5cm}
\noindent{\bf Acknowledgements}\\

We thank E.A. Ivanov, A.P. Nersessian and especially M.A. Olshanetsky  for clarifying  discussions.
This work was partially supported by INTAS grant No 00--254 (S.B., A.G. and S.K.), European Community's Human Potential
Programme contract HPRN-CT-2000-00131 (S.B.) and NATO Collaborative Linkage Grant PST.CLG.979389
(S.B. and A.G.). The work of A.G. has been supported by the President of Russian Federation, grant MD-252.2003.02 and the
Ministry of Education of Russian Federation, grant E02-2.0-7.
S.K. acknowledges the support of DFG, grant No.436
RUS 113/669 and RFBR-DFG 02-02-04002.

\end{document}